\documentstyle[12pt,aasms4]{article}
 
\def\sun{_\odot}

\begin{document}
 
\title{New Observations of Extra-Disk Molecular Gas\\in Interacting
Galaxy Systems,\\
Including a Two-Component System in Stephan's Quintet} 
 
\author{Beverly J. Smith}
\affil{Department of Physics and Astronomy, East Tennessee State
University, Box 70652, Johnson City TN  37604}
\author{Curtis Struck}
\affil{Department of Physics and Astronomy, Iowa State University,
Ames IA  50012}

\affil{}

\begin{abstract}

We present new CO (1 $-$ 0) observations of eleven extragalactic
tails and bridges in nine interacting galaxy systems,
almost doubling the number of such features with sensitive CO
measurements.
Eight of these eleven features were undetected in CO to very low CO/HI
limits, with the most extreme case being the NGC 7714/5 bridge.  This bridge
contains luminous H~II regions and 
has a very high HI column density (1.6 $\times$ 10$^{21}$
cm$^{-2}$ in the 55$''$ CO beam), yet was undetected in CO
to rms T$_R$$^*$ = 2.4 mK.  The HI column density
is higher than standard H$_2$ and CO 
self-shielding limits for solar-metallicity gas, 
suggesting that the gas in this bridge is 
metal-poor and has an enhanced
N$_{H_2}$/I$_{CO}$ 
ratio
compared to the Galactic value. 
Only one of the eleven features in our sample was unambiguously detected in
CO, a luminous HI-rich star formation region near an optical
tail in the compact group Stephan's Quintet.
We detect CO at two widely separated velocities
in this feature, at $\sim$6000 km s$^{-1}$ and $\sim$6700 km
s$^{-1}$.  Both of these components have HI and H$\alpha$ counterparts.
These velocities correspond to those of galaxies
in the group, suggesting that this gas is material
that has been removed from two galaxies in the group.
The CO/HI/H$\alpha$ ratios for both components are similar to
global values for spiral galaxies.

\end{abstract}


\section{Introduction}

Large amounts of interstellar gas can be removed from the main disks of spiral
galaxies by four main processes: tides due to
the gravitational force of a companion, ram pressure stripping during
a near-head-on collision between gas-rich galaxies, 
ram pressure stripping by intracluster gas,
and galactic winds driven by supernovae.   
By ejecting processed gas into intergalactic space,
these mechanisms contribute
to the metal-enrichment of the intergalactic medium.
In many cases, which of these four processes is active in 
a given galaxy system can be determined
from the optical, radio, and/or X-ray morphology.
The signature
of a tidal encounter between two galaxies is the presence
of long stellar and/or HI tails and bridges (e.g., Toomre $\&$ Toomre 1972), 
while head-on collisions between gas-rich galaxies
can produce ring galaxies (e.g., Lynds $\&$ Toomre 1976;
Theys $\&$ Spiegel 1977) as well as
gaseous bridges between the galaxies 
(Struck 1997).
Ram pressure stripping by intracluster gas leads to 
an HI deficiency (e.g., Giovanelli $\&$ Haynes 1983), 
truncation of the outer HI disk of a galaxy 
(Cayette et al. 1990), and in some cases, bending of the HI disk
(Kenney $\&$ Koopmann 1999), but not usually removal of
large quantities of molecular gas (Kenney $\&$ Young
1989).
Galactic winds 
are identified by extended extra-disk ionized gas without stellar
counterparts (e.g., Rand, Kulkarni, $\&$ Hester 1990).

Star formation sometimes occurs in gas clouds
far removed from the main disks of galaxies.
Of the four processes that can remove gas from disks, 
the best-known to trigger
the formation of young stars in the stripped gas is tidal forces:
numerous examples of luminous H~II regions in tidal
features have been found
(e.g., Schweizer 1978; Mirabel et al. 1991, 1992).   
Young stars have also been found in extra-disk gas clouds thought
to have been stripped during head-on collisions (Smith et al. 1999)
and interstellar-intracluster encounters (Xu et al. 1999).

In order to better understand the processes that lead to the removal
of gas from galaxies and the triggering of star formation in this gas,
it is important to make a complete inventory of the gas in these
features.  This means not just 
the atomic gas, which has been surveyed in a large number of
extragalactic tails and bridges 
(e.g.,
Haynes et al. 1984; Smith 1991;
Hibbard $\&$ van Gorkom 1996), but also the 
molecular gas.
Molecular gas has proved elusive in classical tidal tails;
in our earlier CO survey of six tidal tails 
(Smith $\&$ Higdon 1994), no CO
was found to very low levels, while
only very low mass concentrations of molecular gas ($\le$10$^7$ M$\sun$)
were found
in the tidal features of the nearby interacting system M81/M82/NGC 3077
(Brouillet, Henkel, $\&$ Baudry 1992; 
Walter $\&$ Heithausen 1999).

In contrast to tidal features, gas removed from galaxy disks
by ram pressure during head-on collisions between
two gas-rich galaxies may be richer in CO.
The proto-type of this class of object is the 
CO-rich gas concentration found
outside the disk of the Virgo Cluster galaxy NGC 4438
by
Combes et al. (1988).
If the standard Galactic 
N$_{H_2}$/I$_{CO}$ 
ratio
holds in this feature, it contains M$_{H_2}$ $\sim$ 10$^9$ M$\sun$.
The ring-like morphology of NGC 4438 and the
proximity of the companion galaxy NGC 4435 suggests that 
this clump was removed
during a head-on collision between the two galaxies (Kenney et al. 1995).  
Another apparent example of gas ram pressure-stripped
during a galaxy-galaxy collision is the eastern tail of the peculiar
galaxy NGC 2782.  This feature has strong CO emission,
corresponding to
6 $\times$ 10$^8$ M$_{\sun}$, 
assuming the Galactic 
N$_{H_2}$/I$_{CO}$ 
conversion factor
(Smith et al. 1999).
Based on morphological considerations, we surmised
that this feature was created during
a near head-on collision between two galaxies, rather than
a grazing encounter
(Smith 1994).

These results
suggest that features
produced in head-on collisions may 
differ in a fundamental way from classical tidal tails and bridges
formed during grazing encounters: they may be richer in CO. 
During a head-on collision, the material 
pulled out into a tail or bridge may originate 
in the inner disk of one of the galaxies, 
and so may be more metal-rich than gas pulled out from the
outer disk in a more distant encounter.  
The metallicity of the gas may affect the 
N$_{H_2}$/I$_{CO}$ ratio, and therefore the detectability of 
the CO line.  
Both theoretical (e.g., Maloney $\&$ Black 1988) and 
observational (e.g., Wilson 1995; Verter $\&$ Hodge 1995;
Arimoto, Sofue, $\&$ Tsujimoto 1996)
studies show that low metallicities can lead to 
enhanced N$_{H_2}$/I$_{CO}$ ratios compared to the Galactic
value.  
In gas with low abundances and dust content,
ultraviolet radiation penetrates more deeply into a molecular
cloud, causing a larger C$^+$ region relative to the CO core,
increasing the N$_{H_2}$/I$_{CO}$ ratio.

The conclusion that
tidal gas differs from ram pressure stripped gas
is quite uncertain, in part because of the very small
sample size.
We have therefore continued our CO survey of extragalactic tails and
bridges, selecting systems with high HI column densities, strong
star formation rates, and/or ring-like morphologies.
In this paper, we present new CO data for tails and bridges in
nine additional interacting
galaxy systems.  We compare these data with
the previous results discussed above, as well as the new CO data 
discussed in Smith (2000), Braine et al. (2000), and Gao et al. (2000).

Throughout this paper, we assume H$_o$ = 75 km s$^{-1}$ Mpc$^{-1}$.

\section{Observations}

The CO (1 $-$ 0) observations were made 
using the 3mm SIS
receiver on the National Radio Astronomy 
Observatory\footnote{The National Radio Astronomy Observatory
is a facility of the National Science Foundation, operated
under cooperative agreement by Associated Universities, Inc.}
(NRAO) 12m telescope during several observing
runs between 1996 and 2000.
Two 256$\times$2 MHz filterbanks, one for each
polarization, were used for the observations,
providing a total bandpass of 1300 km s$^{-1}$ with
a spectral resolution of 5.2 km s$^{-1}$.
A nutating subreflector with an azimuthal beam throw of 3$'$
was used, taking care to avoid chopping on
other galaxies, tails, or bridges in the system. 
Each scan was 6 minutes
long. The beamsize FWHM is 55$''$ at 115 GHz.
The pointing was checked periodically with
bright continuum sources and was consistent
to 10$''$.  The system temperatures ranged
from 200 to 500 K.  
Calibration was accomplished using an ambient chopper wheel.

A total of 45 positions in nine interacting systems were observed.
These nine systems are listed in Table 1, along with a brief
description of their optical morphologies.  
In Figure 1, we display optical images of these systems, obtained
from the Digitized Sky 
Survey\footnote{The Digitized
Sky Survey was produced at the
Space Telescope Science
Institute under U.S. Government
grant NAG W-2166.
These images are 
digitized versions of 
photographic plates from the 
Second Palomar Observatory
Sky Survey (POSS-II) made by
the California Institute
of Technology with funds from
the National Science Foundation,
the National Geographic Society, the Sloan
Foundation, the Samuel Oshin Foundation, and the Eastman
Kodak Corporation.
} (DSS).
Table 2 lists the observed positions in the galaxies and tails/bridges;
these positions and the CO FWHM beamwidth are marked in Figure 1.
Fourteen of the observed positions were in tails or bridges;
the rest were in the main disks of the galaxies.
A total of eleven tails and bridges were observed.  In three
systems (Arp 144, NGC 2814/20, and NGC 3628), we observed multiple
positions in a single tail or bridge. 

Table 2 also lists the central velocities of the observed bandpasses.
Note that I Zw 192, NGC 3561B, and position A in Stephan's Quintet
were all
observed twice, with two different central
velocities, to increase the observed bandpass; 
these two sets of data were combined.

The results of the CO observations are given in Table 3:
the line fluxes, velocities, widths, and rms noise levels.
The summed spectra for each observed position are displayed in
Figure 2.
Note that the center of NGC 7828 and one of the tail
positions in this system were previously observed by Smith $\&$ Higdon
(1994); the new and old data have been combined.

\section{Results}  

Out of our nine interacting systems, the only unambiguous detection
of CO outside of the main disk of a galaxy is the star formation region
in Stephan's Quintet.    In two of our systems (NGC 3395/6 and
the Taffy Galaxies),
the two galaxies in the pair are separated by only 70$''$ $-$ 90$''$,
and the star formation regions in the bridges are unresolved
from the main disks with our beamsize.  Thus, although
the bridge positions were detected in CO, this may be emission
from the galaxy disk in the 12m beam.  Higher resolution follow-up
observations are needed to confirm the existence of CO 
in these regions.

In Table 4, we
give the molecular gas mass for the galaxy disks,
derived assuming the standard
Galactic 
N$_{H_2}$/I$_{CO}$ 
ratio (2.8 $\times$ 10$^{20}$ cm$^{-2}$/(K km s$^{-1}$);
Bloemen et al. 1986)
and the emission fills the beam ($\eta$$_c$ = 0.82).
For the galaxies where more than one position in the disk was 
detected, the total CO flux from the galaxy was obtained assuming
a Gaussian distribution.
For the observed extra-disk positions in these systems,
the molecular gas mass was derived under the same assumptions,
and are recorded in Table 5.
Although these assumptions may not hold in all cases,
they provide a basis for comparison.
In Table 5, we also provide the HI column densities averaged
over the CO beam, along with the implied M$_{H_2}$/M$_{HI}$
ratios and the beam size $\Theta$, in kpc.

For comparison, in Table 6, we give results for thirteen other tails, bridges,
and extra-disk gas clouds
previously observed in CO.  This includes structures
in NGC 2782 (Smith $\&$ Higdon 1994; Smith et al. 1999),
NGC 4438 (Combes et al. 1988), 
NGC 4676, NGC 7252, and Arp 143 (Smith
$\&$ Higdon 1994), Arp 245 (Braine et al. 2000),
NGC 4410 (Smith 2000),
NGC 3561 (Braine et al. 2000), M81
(Brouillet et al. 1992) and NGC 3077 (Walter $\&$
Heithausen 1999), as well as the tentative (4$\sigma$)
detection of CO in the southern tail of NGC 4038/9 (Gao et al. 2000).
Note that in this paper we present CO data for
the northern tail in the NGC 3561 system, while
data for the southern feature are available from
Braine et al. (2000).
In Tables 5 and 6,
we also give
the H$\alpha$ luminosities for these sources, 
when available, as well as the 
L$_{H\alpha}$/M$_{H_2}$ ratio.

For the sources in Tables 5 and 6
which have HI data available, in Figure 3
we plot the HI column density
in the CO beam
N$_{HI}$ against M$_{H_2}$/M$_{HI}$.  For those
with H$\alpha$ fluxes also available, in Figure 4 we compare
L$_{H\alpha}$/M$_{H_2}$ with M$_{H_2}$/M$_{HI}$.
Figures 3 and 4 show a wide range in  
L$_{H\alpha}$/M$_{H_2}$ and 
M$_{H_2}$/M$_{HI}$ ratios for
the features in our sample.  The most extreme cases
are 1) the NGC 7714/5 bridge, 
with a very high HI column density but very
low CO flux, 
2) the eastern feature of NGC 2782, with a low 
L$_{H\alpha}$/M$_{H_2}$ ratio,
3) the molecular clump near NGC 4438, with a very high implied 
M$_{H_2}$/M$_{HI}$ ratio and a low 
L$_{H\alpha}$/M$_{H_2}$ value, and 
4)
the extra-disk star formation region in Stephan's Quintet, with
a high L$_{H\alpha}$/M$_{H_2}$ ratio.

The global ratios for spiral galaxies are
typically L$_{H\alpha}$/M$_{H_2}$ $\sim$ 0.0025 $-$ 0.1 L$\sun$/M$\sun$
and M$_{H_2}$/M$_{HI}$ $\sim$ 0.2 $-$ 6 (Young et al. 1996).
The star formation region in Stephan's Quintet, the 
Arp 245 feature, the NGC 4038/9 paper, and the upper limits
to the NGC 4676 tail and the NGC 3561 features are
consistent with these global ratios.
In contrast, the features in NGC 2782 and NGC 4438 have very low 
L$_{H\alpha}$/M$_{H_2}$ ratios, and the NGC 7714/5 bridge has a low
implied
M$_{H_2}$/M$_{HI}$ ratio compared to global values for spirals.

Dwarf galaxies typically have higher L$_{H\alpha}$/M$_{H_2}$ ratios and
lower M$_{H_2}$/M$_{HI}$ ratios than normal spirals, if the standard
Galactic conversion factor is applied.
For example, the 14 dwarf galaxies 
in the Sage et al. (1992) sample have 
L$_{H\alpha}$/M$_{H_2}$ $\ge$ 0.04 L$\sun$/M$\sun$
and M$_{H_2}$/M$_{HI}$ $\le$ 0.5, after
converting to the standard Galactic conversion factor, while
all but two of the 25 dwarf galaxies studied by
Israel, Tacconi, $\&$ Baas (1995) 
have M$_{H_2}$/M$_{HI}$ ratios $\le$ 0.02 with this conversion factor.
The inferred molecular gas masses for dwarf galaxies may be low 
because their 
N$_{H_2}$/I$_{CO}$ ratio is enhanced 
due to low metallicities
(Maloney $\&$ Black 1988).
Note, however, that specific locations within dwarf galaxies may
have high implied M$_{H_2}$/M$_{HI}$ ratios; for example, in IC 10,
at positions with N$_H$ $\ge$ 2 $\times$ 10$^{21}$ cm$^{-2}$, 
M$_{H_2}$/M$_{HI}$ $\sim$ 3 (Ohta, Sasaki, $\&$ Saito 1988).

The tails and bridges in Figure 3 that are undetected
in CO 
have upper limits to their M$_{H_2}$/M$_{HI}$ ratios 
consistent with global values for dwarfs.  
The upper limits for the NGC 3561 irregular, the NGC 7714 loop,
the NGC 7715 tail, and the NGC 7714/5 bridge, 
are lower than typical global values for spirals; the rest have less
strict upper limits.

The subset of
these undetected features with H$\alpha$ fluxes available
(NGC 4676,
NGC 7714/5, 
and the northern tail of NGC 3561) have L$_{H\alpha}$/M$_{H_2}$ lower
limits consistent with values for both dwarfs and spirals.
However, the Arp 245 feature, the Stephan's Quintet
source, the NGC 4038/9 tail, the eastern NGC 2782 tail, 
and the NGC 4438 CO clump are brighter
in CO relative to H$\alpha$ than typical dwarf galaxies, while
Stephan's Quintet, the eastern NGC 2782 tail, and the NGC 4438 CO
clump have higher inferred 
M$_{H_2}$/M$_{HI}$ ratios than dwarfs.
The southern NGC 3561 feature has an 
M$_{H_2}$/M$_{HI}$
ratio higher than that usually found in dwarfs, but just 
a lower limit to L$_{H\alpha}$/M$_{H_2}$.

\section{Comments on Individual Galaxies}

\subsection{NGC 7714/5}

The most extreme feature in our sample in terms of
inferred M$_{H_2}$/M$_{HI}$ ratio
is the bridge of NGC 7714/5.
This bridge, which contains luminous H~II regions (Arp 1966;
Bernl\"ohr 1993; Gonz\'alez-Delgado et al. 1995; Smith et al. 1997),
was not detected in CO,
in spite of its high column density 
of atomic hydrogen 
(1.6 $\times$ 10$^{21}$ cm$^{-2}$ in the CO beam; from the data in Smith 
et al. 1997).
At this bridge, our upper limit implies 
N$_{H_2}$ $\le$ 5.0 $\times$ 10$^{19}$ cm$^{-2}$
and M$_{H_2}$/M$_{HI}$ $\le$ 0.063,
more than three times smaller than our upper limit for the 
system with the second lowest M$_{H_2}$/M$_{HI}$ ratio in a tail or
bridge,
the NGC 3561 irregular.
The HI column density
in this bridge is above the H$_2$ and CO self-shielding limits 
for solar metallicity gas
($\sim$5 $\times$ 10$^{20}$ cm$^{-2}$ and 10$^{21}$ cm$^{-2}$,
respectively; Federman et al. 1979; van Dishoeck $\&$ Black 1988).
However, the nucleus of NGC 7714 is known to be low metallicity (French
1980; Garc\'ia-Vargas et al. 1997), so
it is likely that the gas in the bridge would also have low 
abundances, leading to an 
enhanced N$_{H_2}$/I$_{CO}$ ratio.
Thus there may be more molecular gas in this system than implied
by the CO measurements.
We note that the mid-infrared
camera on the Infrared Space Observatory (ISOCAM) did not detect
this bridge (O'Halloran et al. 2000), but it did detect the more distant
extra-disk source in Stephan's Quintet (Xu et al. 1999).
This suggests a smaller warm dust component in the NGC 7714/5 bridge.

The smaller galaxy NGC 7715, the eastern tail of NGC 7715,
and the HI loop north of NGC 7714 are also
weak in CO, with inferred
M$_{H_2}$/M$_{HI}$ ratios $\le$0.16, $\le$0.16, and $\le$0.24, respectively.
These may also have 
enhanced N$_{H_2}$/I$_{CO}$ ratios.  Note that 
NGC 7715, its tail, and the NGC 7714 loop
have not been detected in H$\alpha$ (Smith et al. 1997).
Also note that NGC 7714 and NGC 7715 have absolute blue magnitudes
of $-$19.8 and $-$18.1, respectively (de Vaucouleurs et al. 1991),
compared to $-$17.7 for the Large Magellanic Cloud (de Vaucouleurs
et al. 1991; Sandage, Bell, $\&$ Tripicco 1999).  At absolute
magnitudes fainter than about $-$19, both 
irregular and spiral galaxies typically
have less than solar metallicities (Skillman et al. 1989;
Vila-Costas $\&$ Edmunds 1992; Storchi-Bergmann et al. 1994), thus
NGC 7715 may also be metal-poor.

\subsection{NGC 2782 and NGC 4438}

At the other extreme from the NGC 7714/5 bridge in
terms of CO brightness is the eastern
tail of NGC 2782 and the extra-disk gas cloud near NGC 4438,
with low L$_{H\alpha}$/M$_{H_2}$ ratios compared to
the other features (see Figure 4).  These features have
strong CO emission, but little on-going star formation (Combes
et al. 1988; Kenney et al. 1995; Smith et al. 1999).
None of the features in our new sample
are as rich in CO as the NGC 2782 and NGC 4438 features.
As discussed at length in Smith et al. (1999), the gas in these
features may be metal-rich material removed from the interiors of their disks
by near-head-on collisions, leading to high CO fluxes.
Star formation may be inhibited in these features because of the collision
(Smith et al. 1999).
Gas clouds pushed out of a galactic disk by a high velocity collision
may compress and then adiabatically expand, reducing their self-gravity.
This may decrease the rate of star formation and therefore
the L$_{H\alpha}$/M$_{H_2}$ ratio.

\subsection{Arp 245, NGC 3561, and NGC 4038/9}

In Figure 3 and Tables 5 and 6, 
the structure with the highest HI column density
in the CO beam
is the northern tail of the relatively nearby galaxy Arp 245, which
has
N$_{HI}$ = 2.3 $\times$ 10$^{21}$
cm$^{-2}$ averaged over the 3.5 kpc beam (Duc et al. 2000).
As discussed in Duc et al. (2000), this feature is relatively gas-rich
and CO-rich
(M$_{H_2}$ $\sim$ 1.5 $\times$ 10$^8$ M$\sun$,
M$_{HI}$ $\sim$ 4.8 $\times$ 10$^{8}$ M$\sun$, and
M$_{H_2}$/M$_{HI}$ $\sim$ 0.31
in the CO beam) 
and has
many properties in common with spiral galaxies.
It has an
L$_{H\alpha}$/M$_{H_2}$ ratio similar to spirals, and 
a stellar population dominated by an underlying old population
(Duc et al. 2000).
Furthermore, it has a relatively high extinction (A$_B$ = 2.6,
from the Balmer decrement), a high blue luminosity
(M$_B$ = $-$17.2 and L$_{B}$ = 1.2 $\times$ 10$^9$ L$\sun$,
uncorrected for internal extinction),
and a metallicity of 12 + log(O/H) $\sim$ 8.6 (Duc et al. 2000).
With a modest correction for internal extinction, these values are 
consistent with the metallicity-luminosity relationship
for late-type spirals (Zaritsky, Kennicutt, $\&$ Huchra 1994;
Garnett et al. 1997).  
Thus it is possible that this structure may not be a tidal tail,
but rather a pre-existing edge-on disk galaxy that is interacting with
the other two galaxies in the system.
High resolution kinematic data may be useful in testing this
hypothesis.  In any case, this feature appears to be richer
in CO relative to HI and H$\alpha$ than most
dwarf galaxies.   

The fact that the Arp 245 feature was detected in CO
and many of the other features were not may be due 
to its high HI column density and relatively high
metallicity.  There is some evidence that
N$_{H_2}$/I$_{CO}$ is correlated with
metallicity (Wilson 1995; Verter $\&$ Hodge 1995; Arimoto et al. 1996).
The trends implied in these papers suggest
that at the metallicity of the Arp 245 feature,
the 
N$_{H_2}$/I$_{CO}$ ratio 
is enhanced only slightly, between 1.5 $-$ 3 times bigger than
the Galactic value. Thus, at the high HI column density
of the Arp 245 feature, the CO self-shielding limit is exceeded.
We note that IC 10,
which has a lower oxygen abundance that the Arp 245 feature, shows
strong CO at positions where 
N$_{HI}$ exceeds
10$^{21}$ cm$^{-2}$
(Ohta, Sasaki, $\&$ Saito 1988).
At present, the metal abundances of most of the features
listed in Tables 5 and 6 are unknown.  

The NGC 3561 system (Arp 105; Figure 1d) contains a spiral (NGC 3561A),
an S0 or E galaxy (NGC 3561B), and two tidal features with on-going star
formation, presumably pulled out from the spiral galaxy.
The concentration of gas
and star formation in the northern tail 
is classified as a Magellanic irregular by
Duc $\&$ Mirabel (1994).
The southern feature, which crosses the companion NGC 3561B, is
visible as an optical knot in Figure 1d, and is classified
as a compact dwarf by Duc $\&$ Mirabel (1994).
This southern feature was detected in CO by Braine et al.
(2000), yielding a high M$_{H_2}$/M$_{HI}$ ratio in the CO beam of
$\sim$0.8.  This ratio is higher than that of the Arp 245 feature,
in spite of its more modest HI column density in
the CO beam,
$\sim$ 2 $\times$ 10$^{20}$ cm$^{-2}$ (Duc et al. 1997) and its 
lower oxygen abundance 
(12 + log(O/H) $\sim$ 8.4; Duc $\&$ Mirabel 1994).
This feature
also has a lower blue luminosity 
(M$_{B}$ = $-$16.9; L$_{B}$ = 9.0 $\times$ 10$^8$ L$\sun$, uncorrected
for internal extinction) and a lower extinction (A$_B$ = 1.0)
(Duc et al. 1994) than the Arp 245 feature.

The northern tail of NGC 3561 (the Magellanic irregular)
has a peak HI column density four times higher and
a blue luminosity five times larger than the southern tail
(Duc et al. 1997), and has a higher oxygen abundance
12 + log(O/H) $\sim$ 8.6 (Duc $\&$ Mirabel 1994),
but was undetected in CO, with a lower M$_{H_2}$/M$_{HI}$ ratio
of $\le$0.2.
This difference may be due in part to the difference in beam
sizes of our CO observations and the higher resolution observations
of Braine et al. (2000), which are less affected by beam dilution.
In the interaction scenario for this system presented by
Duc et al. (1997), the spiral galaxy NGC 3561A interacts
with the elliptical NGC 3561B, drawing out a long tail to the
north, and a shorter countertail to the south.  Since both features
may have originated in NGC 3561A, it is reasonable to expect
they would have similar gas properties.
The southern feature may have a relatively small
angular size, high column density clump in which the self-shielding 
limits are exceeded.

If the tentative detection of CO in the NGC 4038/9 tail 
(Gao et al. 2000)
is confirmed
by more sensitive observations, 
it places this tail in the same category as the NGC 3561 and
Arp 245 features: it is richer in CO than dwarf galaxies, relative
to its H$\alpha$ flux.
The metallicity in this feature has been estimated to be
somewhat lower than that in the Arp 245 feature, 
12 + log(O/H) $\sim$ 8.4 (Mirabel et al. 1992).

\subsection{M81 and NGC 3077}

The tidal features of M81 and NGC 3077 have also been detected
in CO (Brouillet et al. 1992; Walter $\&$ Heithausen 1999).
Like those in Arp 245 and NGC 3561, 
they also have high inferred M$_{H_2}$/M$_{HI}$ ratios (Table 6).
However, the inferred
H$_2$ masses in the CO beam are small ($\sim$10$^6$ M$\sun$, using
the standard Galactic 
N$_{H_2}$/I$_{CO}$ 
ratio; 10$^{7}$ M$\sun$,
using the virial theorem) and there is no evidence for 
on-going star formation
in these locations (Brouillet et al. 1992; Henkel et al. 1993;
Walter $\&$ Heithausen 1999).  The M81/M82/NGC 3077 group is very nearby,
so the area subtended by the CO beam ($\sim$360 pc)
is much
smaller than in the other systems listed.
The other features plotted in Figure 3 have physical beam diameters
ranging from 1.8 kpc (NGC 4438) to 43 kpc (I Zw 192).
As with the Arp 245 and NGC 3561 features, the CO self-shielding
threshold may be exceeded locally in these features.  We note
that the HI column density in the CO beam is quite high for the
M81 and NGC 3077 positions,
10$^{21}$ cm$^{-2}$.

A recent larger area CO survey of the NGC 3077 tail (Walters $\&$
Heithausen 2000) shows CO in two clumps covering an area of
about 3$'$ (2.8 kpc).
In this region, the standard Galactic conversion factor gives
M$_{H_2}$ $\sim$ 6 $\times$ 10$^6$ M$\sun$ and an average
H$_2$ column density of $\sim$8 $\times$ 10$^{19}$ cm$^{-2}$.  
As in IC 10,
CO is only detected at HI column densities above $\sim$ 10$^{21}$ cm$^{-2}$.
In this larger area, the inferred 
M$_{H_2}$/M$_{HI}$ ratio is smaller, $\sim$ 0.1, consistent with
the upper limits on the undetected tails and bridges in the sample.

\subsection{Stephan's Quintet}

Of the nine extra-disk locations in our new sample, only one was detected
in CO: the position in 
Stephan's Quintet (HCG 92; Arp 319).
Stephan's Quintet is a compact group of four galaxies with
similiar velocities ($\sim$6000 km s$^{-1}$;
NGC 7317, NGC 7318A/B, and NGC 7319), and a probable foreground
galaxy (NGC 7320) with a much lower velocity ($\sim$800 km s$^{-1}$).
The outlying galaxy NGC 7320C also has a velocity of $\sim$6000 km s$^{-1}$,
and so probably also belongs to this group.
Strong gravitational interactions are clearly present
in this group (Figure 1g).
A long optical tail
stretches more than 1$'$ (26 kpc) to the southeast of NGC 7319; 
this tail is also detected in HI (Shostak
et al. 1984; Verdes-Montenegro et al. 2000).
Two more optical
tails are seen extending to the north of the 
close pair NGC 7318A/B 
(see Figure 1g).  Along the easternmost of these features,
X-rays (Pietsch et al. 1997) and radio continuum emission (van der Hulst
$\&$ Rots 1981) are visible, suggesting a shock front.
At the location where the two optical tails of NGC 7318A/B intersect,
a large concentration of ionized gas has been detected (Moles,
Sulentic, $\&$ M\'arquez 1997; Ohyama
et al. 1998; Xu
et al. 1999).  This source is also present in the optical
DSS image (Figure 1g), in the mid-infrared map of Xu et al. (1999),
and in HI (Shostak, Sullivan, $\&$ Allen 1984; Verdes-Montenegro
et al. 2000).  
Optical spectroscopy shows that
the ionization mechanism for this
source is young stars, while
spectral signatures of shocks are present to the south, along the 
radio continuum and X-ray ridge 
(Moles, M\'arquez, $\&$ Sulentic 1998).

At the location
of the extra-disk star formation region in Stephan's Quintet,
we have detected two CO components,
at 6000 km s$^{-1}$ 
and 6700 km s$^{-1}$, at the 5$\sigma$
and 6$\sigma$ significance level, respectively.
These two components have similar CO fluxes (Table 3).
The large velocity separation of these two features argues
against the idea that they are caused by the double-horned
signature of a rotating disk, but are instead two distinct
gas clouds.
These two components each contain 1.0 $\times$ 10$^9$ M$\sun$ of molecular
gas, if the standard Galactic 
N$_{H_2}$/I$_{CO}$ 
ratio holds.
One of these components (6000 km s$^{-1}$) was detected at a 4$\sigma$
level in the
CO interferometer data of Gao $\&$ Xu (2000).
We have also detected CO in NGC 7319 and NGC 7318B.
NGC 7319
had previously been seen in CO (Verdes-Montenegro
et al. 1998; Yun et al. 1997), however, the weaker CO
flux from 7318B was not seen
in those less sensitive studies.
The CO detection in NGC 7318B is at 6000 km s$^{-1}$,
a somewhat higher velocity than the 
optical velocity of this galaxy (5765 km s$^{-1}$;
Moles et al. 1998).

Both of the CO components in the extra-disk star formation
region have HI counterparts (Shostak et al. 1984).
They also both
appear to have on-going star formation.
Moles et al. (1998) obtained optical spectra of three H~II regions
in this vicinity; two have velocities of $\sim$6020 km s$^{-1}$
and one has a velocity of 6680 km s$^{-1}$.
Two H$\alpha$ velocity components are also indicated by the narrowband
imaging of Xu et al. (1999).

The H$\alpha$ luminosity, molecular
gas mass, and HI mass for the extra-disk
source in Stephan's Quintet
are similar to those of many normal spiral galaxies 
(Young et al. 1996), but
unlike those of typical dwarf galaxies (Sage et al. 1992).
The implied M$_{H_2}$/M$_{HI}$ ratios for both velocity components
in this region
are very high, $\sim$0.65, higher than values typically
found for dwarf galaxies (Sage et al. 1992).
The L$_{H\alpha}$/M$_{H_2}$
ratio for the Stephan's Quintet source, 0.016 L$\sun$/M$\sun$,
is consistent with
global values for spirals (Young et al. 1996), as 
well as the values found for  
the NGC 4676 tail, the Arp 245 feature, and the southern tail
of NGC 3561, but is higher than the values seen for the
eastern tail of NGC 2782 and the NGC 4438 clump.
In Smith et al. (1999), we suggested that star formation
was inhibited in the eastern tail of NGC 2782 and the
NGC 4438 CO clump because of strong shocks sustained during
a head-on collision between two gas-rich
galaxies.  If the extra-disk cloud in Stephan's Quintet
was formed in
this way, star formation is clearly not inhibited.

To explain the peculiar morphology of
Stephan's Quintet, a number of different interaction scenarios
have been suggested.  Some of these scenarios involve
ram pressure stripping during a head-on
collision between two galaxies.
For example, Peterson $\&$ Shostak (1980) suggest that
a head-on collision between NGC 7318B and NGC 7319
removed the gas and caused the radio continuum ridge between them.
A different scenario was suggested by
Shostak et al. (1984): a prograde tidal 
encounter of NGC 7320C
and NGC 7319 occurred, pulling gas out of these galaxies, creating
the long
tail south of NGC 7319.  This was
followed by a high-speed collision
between NGC 7318A and NGC 7318B.
Other scenarios involve 
a collision between an `intruder galaxy'
(usually NGC 7318B) and intragalactic gas 
that had previously been tidally-stripped or ram pressure-stripped
in an earlier encounter between two different galaxies in the group
(van der Hulst $\&$ Rots 1981;
Moles et al. 1997).
Moles et al. (1997) suggest that 
a direct collision between
the outlying galaxy NGC 7320C and NGC 7319 occurred in
the recent past (10$^8$ years ago), 
removing a large quantity of gas from these
galaxies.  After this event,
NGC 7318B entered the group at high velocity, colliding with
this stripped gas and triggering star formation.

Our new data shed some light on the question of the origin of
the extra-disk gas in Stephan's Quintet.
Our detection of
two CO-rich components in the extra-disk star formation
region shows that the star
formation in this feature was triggered by a
collision involving relatively high metallicity, molecule-rich gas 
concentrations, rather than metal-poor intracluster medium.
This does not rule out ram pressure 
stripping of NGC 7318B by an intergalactic
medium, but it does require that this intergalactic medium be 
relatively metal-rich gas: gas previously removed from a galaxy.
How this gas was originally removed from the disk is
unclear: as shown in Arp 245 and NGC 3561,
structures apparently created by tidal forces can
be relatively CO-rich.
Thus this intergalactic gas may have been removed either
by tidal forces, or via ram pressure stripping.

The 6700 km s$^{-1}$ extra-disk component may
have originally come from NGC 7319 or NGC 7318A, which
have similar optical velocities (6650 km s$^{-1}$ and 6620 km s$^{-1}$,
respectively; Moles et al. 1998).
The most likely candidate of these two is the barred spiral NGC 7319,
which shows a surprising
lack of H~II regions 
(Arp 1973), a long optical tail,
and a very offset CO distribution (Yun et al. 1997).  
NGC 7318A also has an optical tail, however,
this may have been created by another interaction.

The 6000 km s$^{-1}$
extra-disk gas may have originated in 
either the outlying galaxy NGC 7320C or in NGC 7318B.
NGC 7320C has an optical radial velocity
of 6000 $\pm$ 150 km s$^{-1}$ (Lynds 1972), consistent with
this velocity.
Previous arguments for the involvement of NGC 7320C
(Moles et al. 1997)
cited the long tail south of NGC 7319 that points
towards NGC 7320C. 
Alternatively, this component of the 
extra-disk gas may have come from NGC 7318B.
As noted above, 
our CO observations 
show a faint CO line in NGC 7318B at this velocity, redshifted 300 km s$^{-1}$ 
from the optical velocity.
This detection corresponds to
a molecular gas mass of 1.7 $\times$ 10$^9$ M$\sun$, assuming the standard
N$_{H_2}$/I$_{CO}$ 
conversion factor.

We therefore suggest that the extra-disk gas undergoing star
formation in Stephan's Quintet originated in NGC 7318B
and NGC 7319.
At the present time it is unclear whether a single collision
between NGC 7318B and NGC 7319 occurred, ram pressure stripping 
the gas and creating the shock front between the two galaxies,
or whether gas had been removed from NGC 7319 earlier, by
an earlier encounter with another galaxy,
and then this gas was impacted
by NGC 7318B.  

\section{Conclusions}

We have obtained new CO (1 $-$ 0) observations of eleven extragalactic
tails and bridges,
and compared these measurements with 
previously-published data for another thirteen features.
Of these 24 structures,
four have inferred 
M$_{H_2}$/M$_{HI}$ upper limits
(assuming the Galactic 
N$_{H_2}$/I$_{CO}$ conversion factor) less than global values for
spiral galaxies, consistent with those found for irregular galaxies.
Three of these four CO-poor features are in the NGC 7714/5 system;
the most extreme case is the star forming bridge,
which has a very high HI column density (1.6 $\times$ 10$^{21}$
cm$^{-2}$ in the 55$''$ CO beam), yet was undetected in CO,
giving an inferred
M$_{H_2}$/M$_{HI}$ upper limit of $\le$0.063.
These features may have enhanced 
N$_{H_2}$/I$_{CO}$
ratios compared to the Galactic value due to low metallicities.

Of the 24 structures in our combined sample, eight have been detected in CO.
Three of these eight (the Stephan's Quintet extra-disk source,
the NGC 4038/9 tail, and
the Arp 245 tail) have
inferred
efficiencies of star formation 
(L$_{H\alpha}$/M$_{H_2}$) consistent with global values
for spiral galaxies.  In contrast, the 
L$_{H\alpha}$/M$_{H_2}$
ratios for
the eastern tail of NGC 2782 and the 
NGC 4438 clump are lower than typical for disk galaxies, suggesting
inhibited star formation.
In none of the features were our CO limits low enough to
conclusively show 
L$_{H\alpha}$/M$_{H_2}$ ratios as high as those found in 
dwarf galaxies.

Of the eight detected features, two (NGC 2782
and NGC 4438) have optical morphologies
indicating head-on 
collisions.  The Stephan's Quintet feature may also have
been formed in this manner.  In these systems, we suggest
that metal-rich gas 
with an approximately Galactic
N$_{H_2}$/I$_{CO}$ ratio 
was pulled from the inner parts
of the galaxies.

A few recent studies have reported the
detection of CO in a number of 
classical tidal features (e.g., NGC 4038/9, Arp 245,
and NGC 3561).
In these structures, the HI column density may be locally
high enough that the CO self-shielding limit is exceeded,
in spite of modestly enhanced 
N$_{H_2}$/I$_{CO}$ ratios.
There is a slight tendency for 
these 
tidal features to have lower 
M$_{H_2}$/M$_{HI}$ ratios
and higher
L$_{H\alpha}$/M$_{H_2}$ ratios than ram pressure stripped features
discussed above,
but there is a large amount of scatter and the sample size is small.
More observations are needed to determine whether there
is truly a statistical difference in the gaseous and 
star formation properties 
of tidal and ram pressure stripped features.

\vskip 0.1in

We are grateful for the help of the NRAO 12m telescope operators
and staff in obtaining these data.   
This research
has made use of the NASA/IPAC Extragalactic Database (NED)
which is operated by the Jet Propulsion Laboratory under contract
with NASA.  We are pleased to acknowledge partial funding for
this project from a NASA grant administrated by the American
Astronomical Society and from NSF grant
INT-9908542.

\vfill
\eject

{\bf Captions }

Figure 1.  Optical images of the nine interacting galaxy
systems in our sample, with the locations and FWHM beamsizes
of the CO beams
marked.
All images are from the Digitized Sky Survey (DSS), except
when noted.
a) Arp 144 (NGC 7828/9).
There is an HI tail without a known optical counterpart to the 
southeast of the optical pair (Higdon 1988).
b) NGC 2814/20/MK 208.
c) NGC 3395/6.
d) Arp 105 (NGC 3561).
This image has been smoothed by a 3$''$ FWHM
Gaussian to enhance low surface brightness features.
e) The brightest galaxy in the Leo Triplet, NGC 3628.
A long HI tail extends to the east of NGC 3628
(Rots
1978; Haynes, Giovanelli, $\&$ Roberts 1979);
this tail has a faint optical counterpart not visible
in the DSS image shown here
(Kormendy $\&$ Bahcall 1974).
The two marked locations are in this tail.
We also observed the center of NGC 3628.
f) I Zw 192.
g) Stephan's Quintet.
h) A narrowband red continuum image of 
NGC 7714/5 (Arp 284), from Smith et al. (1997).
i) UGC 12914/5 (Taffy Galaxies).

Figure 2. The summed CO (1 $-$ 0) scans for the
observed positions,
after the spectra have been smoothed by a 36 km s$^{-1}$ boxcar
and then resampled at 21 km s$^{-1}$ spacing.

Figure 3. A comparison of N$_{HI}$ with M$_{H_2}$/M$_{HI}$
for the extra-disk positions in Tables 5 and 6 with both HI and CO data.
For comparison purposes, the standard Galactic 
N$_{H_2}$/I$_{CO}$ 
ratio is used to calculate the molecular gas masses.  As discussed
in the text, this ratio may not hold in all cases.
The unlabeled points 
near M$_{H_2}$/M$_{HI}$ $\sim$ 0.8 $-$ 0.9 and
N$_{HI}$ $\sim$ 10$^{20}$ cm$^{-2}$
are the data
for the Arp 143 tail, the two positions in the 
Arp 144 tail, the NGC 2782 western tail,
and the NGC 4410 tail (see Tables 5 and 6).

Figure 4. The M$_{H_2}$/M$_{HI}$ ratio plotted against
the L$_{H\alpha}$/M$_{H_2}$ ratio for the extra-disk sources
with HI, CO, and H$\alpha$ measurements available.

\eject

\begin{center}
   {\bf TABLE 1}\\
   INTERACTING GALAXY SYSTEMS IN SAMPLE$^a$\\ [12pt]
   \begin{tabular}{cccrrrrrcrcccclclcrcccc} \hline
       \multicolumn{1}{c}{\bf System Name}&
\multicolumn{1}{c}{\bf Galaxies/Tails/Bridges}
&\multicolumn{1}{c}{\bf Optical and HI Morphology}\\
Arp 144&NGC 7828&Ring Galaxy$^b$\\
&NGC 7829&Spheroidal$^b$\\
&HI Plume$^c$&No Detected Optical Counterpart$^c$\\
NGC 2814/2820 Group&NGC 2814&Sb\\
&NGC 2820&SB(s)c pec\\
&MK 208&I0 Pec; near NGC 2814/20 Bridge\\
&NGC 2814/20 Bridge&Visible in Radio Continuum$^d$\\
NGC 3395/6&NGC 3395&SAB(rs)cd pec\\
&NGC 3396&IBm pec\\
&NGC 3395/6 Bridge&Visible in Radio Continuum$^e$\\
Arp 105&NGC 3561A&SA(r)a Pec\\
&NGC 3561B&S0 Pec\\
&NGC 3561 Irregular$^f$&Connected by Stellar Bridge to NGC 3561A\\
&NGC 3561 Compact Dwarf$^f$&Immediately South of NGC 3561B\\
Leo Triplet&NGC 3627&SAB(s)b\\
&NGC 3623&SAB(rs)a\\
&NGC 3628&Sb Pec\\
&NGC 3628 Tail&Visible in Optical and HI Maps$^g$\\
I Zw 192 System&I Zw 192&Ring or Loop-like Optical Structure$^h$\\
&I Zw 192 Companion&Compact Galaxy Connected by Bridge$^h$\\
&I Zw 192 Plume&Plume Extending from Companion$^h$\\
\\
   \end{tabular}
\end{center}

\vfill
\eject

\begin{center}
   {\bf TABLE 1 (Continued)}\\
   INTERACTING GALAXY SYSTEMS IN SAMPLE$^a$\\ [12pt]
   \begin{tabular}{cccrrrrrcrcccclclcrcccc} \hline
       \multicolumn{1}{c}{\bf System Name}&
\multicolumn{1}{c}{\bf Galaxies/Tails/Bridges}
&\multicolumn{1}{c}{\bf Optical and HI Morphology}\\
Stephan's Quintet&NGC 7317&E4\\
&NGC 7318A&E2 pec\\
&NGC 7318B&SB(s)bc pec\\
&NGC 7319&SB(s)bc pec\\
&NGC 7320&SA(s)d; Foreground Galaxy\\
&NGC 7320C&(R)SAB(s)0/a\\
&Star Formation Region A&North of NGC 7318B, Near Optical Tails\\
Taffy Galaxies&UGC 12914&(R)S(r)cd pec\\
&UGC 12915&Sc\\
&UGC 12914/5 Bridge&Visible in Radio Continuum and HI$^i$\\
Arp 284&NGC 7714&SB(s)b pec\\
&NGC 7715&Im pec\\
&NGC 7714/5 Bridge&HI Offset from Optical Bridge$^j$\\
&NGC 7714 Loop&Visible in Optical and HI Maps$^j$\\
&NGC 7715 Tail&Visible in Optical and HI Maps$^j$\\
\\

   \end{tabular}
\end{center}

$^{a}${All information from the NASA Extragalactic Database (NED), unless
otherwise noted.}
$^{b}${Freeman $\&$ de Vaucouleurs 1974.}
$^{c}${Higdon 1988; Figure 1a.}
$^{d}${Bosma et al. 1980; van der Hulst $\&$ Hummel 1985.}
$^{e}${Huang et al. 1994.}
$^{f}${Duc $\&$ Mirabel 1994.}
$^{g}${Kormendy $\&$ Bahcall 1974; Rots
1978; Haynes, Giovanelli, $\&$ Roberts 1979.}
$^{h}${Smith 1989; Figure 1f.}
$^{i}${Condon et al. 1993.}
$^{j}${Smith et al. 1997.}

\vfill
\eject

\begin{center}
   {\bf TABLE 2}\\
   CO (1-0) OBSERVATIONS\\ [12pt]
   \begin{tabular}{crrrrrrrrcccclclcrcccc} \hline
       \multicolumn{1}{c}{Name}&
\multicolumn{6}{c}{Position Observed}
&\multicolumn{1}{c}{Central Velocity}\\
       \multicolumn{1}{c}{}&
\multicolumn{3}{c}{R.A. (1950)}&
\multicolumn{3}{c}{Dec. (1950)}&
\multicolumn{1}{c}{(km s$^{-1}$)}\\
\hline
NGC 7828&0&3&53.7&$-$13&41&40.0&5770\\
Arp 144 Tail $\#1$&0&4&4.2&$-$13&43&57.7&5700\\
Arp 144 Tail $\#2$&0&4&10.2&$-$13&45&16.0&5700\\
NGC 2814&9&17&9.2&64&27&50.0&1634\\
NGC 2814/20 Bridge&9&17&15.5&64&27&23.0&1580\\
MK 208&9&17&26.9&64&27&7.0&1580\\
NGC 2820&9&17&43.7&64&28&16.0&1580\\
NGC 2820 NE1&9&17&47.0&64&28&28.9&1580\\
NGC 2820 NE2&9&17&50.3&64&28&41.8&1580\\
NGC 2820 NE3&9&17&53.6&64&28&54.6&1580\\
NGC 2820 NE4&9&17&56.9&64&29&7.5&1580\\
NGC 3395&10&47&2.7&33&14&44.0&1620\\
NGC 3395/6 Bridge&10&47&6.0&33&15&5.8&1620\\
NGC 3396&10&47&8.9&33&15&18.0&1620\\
NGC 3561A&11&8&31.3&28&59&0.9&8810\\
NGC 3561 Irregular&11&8&31.4&29&2&12.1&8670\\
NGC 3561B&11&8&31.5&28&58&5.5&8500, 8700\\
NGC 3628&11&17&40.3&13&51&46.0&880\\
NGC 3628 Tail $\#1$&11&19&46.3&13&54&57.0&880\\
NGC 3628 Tail $\#2$&11&20&38.0&13&59&0.0&880\\
\\
   \end{tabular}
\end{center}

\vfill
\eject

\begin{center}
   {\bf TABLE 2 (Continued)}\\
   CO (1-0) OBSERVATIONS\\ [12pt]
   \begin{tabular}{crrrrrrrrcccclclcrcccc} \hline
       \multicolumn{1}{c}{Name}&
\multicolumn{6}{c}{Position Observed}
&\multicolumn{1}{c}{Central Velocity}\\
       \multicolumn{1}{c}{}&
\multicolumn{3}{c}{R.A. (1950)}&
\multicolumn{3}{c}{Dec. (1950)}&
\multicolumn{1}{c}{(km s$^{-1}$)}\\
\hline
I Zw 192&17&39&14.4&38&45&21.0&12100, 12300\\
I Zw 192 Companion&17&39&18.4&38&46&12.0&12300\\
I Zw 192 Plume&17&39&19.6&38&46&37.0&12300\\
NGC 7318A&22&33&39.3&33&42&22.2&6630\\
NGC 7318B&22&33&40.9&33&42&24.1&5774\\
Stephan's Quintet A&22&33&41.2&33&43&21.5&5774, 6630\\
NGC 7319&22&33&46.0&33&42&59.4&6764\\
NGC 7714 West&23&33&38.9&1&52&42.0&2800\\
NGC 7714 South&23&33&40.6&1&52&17.0&2800\\
NGC 7714 Center&23&33&40.6&1&52&42.0&2800\\
NGC 7714 North&23&33&40.6&1&53&7.0&2800\\
NGC 7714 Loop&23&33&40.6&1&53&32.0&2800\\
NGC 7714 East&23&33&42.3&1&52&42.0&2800\\
NGC 7714/5 Bridge&23&33&44.9&1&52&50.0&2800\\
NGC 7715 Center&23&33&48.3&1&52&48.0&2800\\
NGC 7715 Tail&23&33&51.1&1&53&12.0&2800\\
UGC 12914 SW2&23&59&0.4&23&11&44.0&4371\\
UGC 12914 NW2&23&59&1.6&23&13&17.0&4371\\
UGC 12914 SW1&23&59&2.4&23&12&3.0&4371\\
UGC 12914 NW1&23&59&3.0&23&12&50.0&4371\\
\\
   \end{tabular}
\end{center}

\vfill
\eject

\begin{center}
   {\bf TABLE 2 (Continued)}\\
   CO (1-0) OBSERVATIONS\\ [12pt]
   \begin{tabular}{crrrrrrrrcccclclcrcccc} \hline
       \multicolumn{1}{c}{Name}&
\multicolumn{6}{c}{Position Observed}
&\multicolumn{1}{c}{Central Velocity}\\
       \multicolumn{1}{c}{}&
\multicolumn{3}{c}{R.A. (1950)}&
\multicolumn{3}{c}{Dec. (1950)}&
\multicolumn{1}{c}{(km s$^{-1}$)}\\
\hline
UGC 12914 Center&23&59&4.4&23&12&22.0&4371\\
UGC 12914 SE1&23&59&5.8&23&11&54.0&4371\\
UGC 12914 Bridge&23&59&6.4&23&12&41.0&4371\\
UGC 12914 SE2&23&59&7.2&23&11&26.8&4371\\
UGC 12915 Center&23&59&8.3&23&13&00.0&4371\\
\\
   \end{tabular}
\end{center}

\vfill
\eject

\begin{center}
   {\bf TABLE 3}\\
   CO (1-0) RESULTS \\ [12pt]
   \begin{tabular}{cccrrrrrrcccclclcrcccc} \hline
       \multicolumn{1}{c}{Name}
&\multicolumn{1}{c}{T$_R$$^*$ (rms)$^a$}
&\multicolumn{1}{c}{$I_{CO}$$^b$}
&\multicolumn{1}{c}{Line Velocity}&
\multicolumn{1}{c}{$\Delta$V$^c$}\\
       \multicolumn{1}{c}{}&
\multicolumn{1}{c}{(mK)}& 
\multicolumn{1}{c}{(K km s$^{-1}$)}& 
\multicolumn{1}{c}{(km s$^{-1}$)}& 
\multicolumn{1}{c}{(km s$^{-1}$)}&\\
\hline
NGC 7828$^d$&2.5&1.16 $\pm$ 0.13&5620&625\\
Arp 144 Tail $\#1$&2.5&$\le$0.13$^e$\\
Arp 144 Tail $\#$2$^d$&2.4&$\le$0.13$^e$\\
NGC 2814&2.0&0.27 $\pm$ 0.07&1710&260\\
NGC 2814/20 Bridge&3.6&$\le$0.59$^f$\\
MK 208&3.0&$\le$0.49$^f$\\
NGC 2820&4.3&1.39 $\pm$ 0.23&1620&570\\
NGC 2820 NE1&5.1&$\le$0.83$^f$\\
NGC 2820 NE2&5.3&$\le$0.87$^f$\\
NGC 2820 NE3&7.4&$\le$1.21$^f$\\
NGC 2820 NE4&5.5&$\le$0.90$^f$\\
NGC 3395&4.4&1.19 $\pm$ 0.17&1600&360\\
NGC 3395/6 Bridge&3.6&0.92 $\pm$ 0.18&1660&420\\
NGC 3396&5.1&0.55 $\pm$ 0.14&1700&150\\
NGC 3561A&4.0&2.12 $\pm$ 0.21&8750&550\\
NGC 3561 Irregular&2.3&$\le$0.20$^g$\\
NGC 3561B&3.8, 4.3, 2.8&$\le$0.19$^g$\\
NGC 3628&4.8&29.58 $\pm$ 0.25&860&400\\
NGC 3628 Tail $\#1$&2.3&$\le$0.13$^h$\\
NGC 3628 Tail $\#2$&2.2&$\le$0.13$^h$\\
\\
   \end{tabular}
\end{center}

\begin{center}
   {\bf TABLE 3}\\
   CO (1-0) RESULTS \\ [12pt]
   \begin{tabular}{cccrrrrrrcccclclcrcccc} \hline
       \multicolumn{1}{c}{Name}
&\multicolumn{1}{c}{T$_R$$^*$ (rms)$^a$}
&\multicolumn{1}{c}{$I_{CO}$$^b$}
&\multicolumn{1}{c}{Line Velocity}&
\multicolumn{1}{c}{$\Delta$V$^c$}\\
       \multicolumn{1}{c}{}&
\multicolumn{1}{c}{(mK)}& 
\multicolumn{1}{c}{(K km s$^{-1}$)}& 
\multicolumn{1}{c}{(km s$^{-1}$)}& 
\multicolumn{1}{c}{(km s$^{-1}$)}&\\
\hline
I Zw 192&2.5, 1.9, 1.3&1.41 $\pm$ 0.07&12130&600\\
I Zw 192 Companion&2.0&$\le$0.34$^i$\\
I Zw 192 Plume&3.1&$\le$0.52$^i$\\
NGC 7318A&12.7&$\le$2.13$^j$\\
NGC 7318B&2.3&0.47 $\pm$ 0.12&6000&400\\
Stephan's Quintet A&1.8, 1.7, 1.0 &0.296 $\pm$ 0.059, 0.286 
$\pm$ 0.048&6000, 6700&140, 180\\
NGC 7319&2.7&1.90 $\pm$ 0.15&6730&470\\
NGC 7714 West&4.1&0.65 $\pm$ 0.14&2800&230\\
NGC 7714 South&3.0&0.50 $\pm$ 0.14&2780&390\\
NGC 7714 Center&2.6&1.43 $\pm$ 0.08&2800&200\\
NGC 7714 North&3.9&0.84 $\pm$ 0.13&2800&220\\
NGC 7714 Loop&4.4&$\le$0.33$^k$\\
NGC 7714 East&3.1&1.21 $\pm$ 0.11&2850&250\\
NGC 7714/5 Bridge&2.4&$\le$0.15$^k$&\\
NGC 7715 Center&2.7&$\le$0.21$^k$&\\
NGC 7715 Tail&3.1&$\le$0.16$^k$\\ 
UGC 12914 SW2&11.4&$\le$1.91$^l$\\
UGC 12914 NW2&13.5&$\le$2.26$^l$\\
UGC 12914 SW1&8.2&$\le$1.37$^l$\\
UGC 12914 NW1&13.1&4.89 $\pm$ 0.75&4200&620\\ 
\\
   \end{tabular}
\end{center}

\begin{center}
   {\bf TABLE 3}\\
   CO (1-0) RESULTS \\ [12pt]
   \begin{tabular}{cccrrrrrrcccclclcrcccc} \hline
       \multicolumn{1}{c}{Name}
&\multicolumn{1}{c}{T$_R$$^*$ (rms)$^a$}
&\multicolumn{1}{c}{$I_{CO}$$^b$}
&\multicolumn{1}{c}{Line Velocity}&
\multicolumn{1}{c}{$\Delta$V$^c$}\\
       \multicolumn{1}{c}{}&
\multicolumn{1}{c}{(mK)}& 
\multicolumn{1}{c}{(K km s$^{-1}$)}& 
\multicolumn{1}{c}{(km s$^{-1}$)}& 
\multicolumn{1}{c}{(km s$^{-1}$)}&\\
\hline
UGC 12914 Center&6.1&6.50 $\pm$ 0.29&4330&600\\
UGC 12914 SE1&8.2&4.10 $\pm$ 0.33&4550&310\\
UGC 12914 Bridge&8.1&12.63 $\pm$0.53&4500&830\\
UGC 12914 SE2&9.7&$\le$1.63$^l$\\
UGC 12915 Center&8.7&12.07 $\pm$ 0.43&4460&730\\
\\
   \end{tabular}
\end{center}

$^a$As noted in Table 2 and in the text, at three positions
two sets of observations were made at two different central velocities,
in order to increase the observed bandpass.
In these cases, the rms values listed correspond to the noise
levels for the lower velocity spectrum, the higher velocity spectrum,
and the combined overlap region, respectively.
Note that for Stephan's Quintet A, two lines were detected, at two different
velocities.
$^b$Statistical uncertainties only.
$^c$Full width zero maximum (FWZM) line widths.
$^d$Combined with data from Smith $\&$ Higdon 1994.
$^e$Using $\Delta$v = 60 km s$^{-1}$, from the HI data of Higdon 1988.
$^f$Using the CO line width of NGC 2820 of 570 km s$^{-1}$.
$^g$Using the HI line widths of 98 km s$^{-1}$ for
NGC 3561B and 161 km s$^{-1}$ for the NGC 3561 dwarf,
from Duc et al. 1997.
$^h$Using the HI line width of 70 km s$^{-1}$ from Rots 1978.
$^i$Using the CO line width of I Zw 192, 600 km s$^{-1}$.
$^j$Assuming a line width of $\le$600 km s$^{-1}$.
$^k$Using the HI line widths of 80 km s$^{-1}$ for the bridge, 130 
km s$^{-1}$ for NGC 7715, 60 km s$^{-1}$ for the eastern tail
of NGC 7715, and 120 km s$^{-1}$ for the northern loop.
These widths were obtained from the combined
HI dataset of Smith $\&$ Wallin 1992 and Smith et al. 1997.
$^l$Using the CO line width of UGC 12914, 600 km s$^{-1}$.

\vfill
\eject

\begin{center}
   {\bf TABLE 4}\\
PARAMETERS OF THE MAIN DISKS OF THE SAMPLE GALAXIES\\ [12pt]
   \begin{tabular}{crrrrrrrrcccclclcrcccc} \hline
       \multicolumn{1}{c}{Name}&
\multicolumn{1}{c}{M(H$_2$)$^a$}\\
       \multicolumn{1}{c}{}&
\multicolumn{1}{c}{(M$_{\sun}$)}\\
\hline
NGC 7828&3.0 $\times$ 10$^9$\\
NGC 2820&3.1 $\times$ 10$^8$\\
NGC 2814&6.0 $\times$ 10$^7$\\
NGC 3395&2.4 $\times$ 10$^8$\\
NGC 3396&1.1 $\times$ 10$^8$\\
NGC 3561A&1.3 $\times$ 10$^{10}$\\
NGC 3561B&$\le$1.2 $\times$ 10$^9$\\
NGC 3628&1.6 $\times$ 10$^9$\\
I Zw 192&1.7 $\times$ 10$^{10}$\\
I Zw 192 Companion&$\le$4.1 $\times$ 10$^9$\\
NGC 7318A&$\le$7.6 $\times$ 10$^9$\\
NGC 7318B&1.7 $\times$ 10$^9$\\
NGC 7319&6.7 $\times$ 10$^9$\\
NGC 7714&2.2 $\times$ 10$^9$\\
NGC 7715&$\le$1.3 $\times$ 10$^8$\\
UGC 12914&1.7 $\times$ 10$^{10}$\\
UGC 12915&1.9 $\times$ 10$^{10}$\\
\\
   \end{tabular}
\end{center}

$^a$Calculated assuming the standard Galactic 
N$_{H_2}$/I$_{CO}$ 
ratio
(M$_{H_2}$ = 1.1 $\times$ 10$^4$ D$^2$ $\int$S$_V$dV, where
D is the distance in Mpc; Bloemen et al. 1986), 
H$_o$ = 75 km s$^{-1}$ Mpc$^{-1}$, and assuming 34 Jy/K for the 12m
telescope and the source fills the beam
($\eta$$_c$ = 0.82). 
For the two sources with more than one position detected in the
disk, the data were fit to a Gaussian distribution as in Young et al. 1995.

\eject
\vfill

\begin{center}
   {\bf TABLE 5}\\
PARAMETERS OF THE EXTRA-DISK REGIONS$^a$\\ [12pt]
   \begin{tabular}{crrrrrrrrcccclclcrcccc} \hline
       \multicolumn{1}{c}{Name}&
\multicolumn{1}{c}{$\Theta$}&
\multicolumn{1}{c}{M$_{H_2}$$^b$}&
\multicolumn{1}{c}{N$_{H_2}$$^b$}&
\multicolumn{1}{c}{N$_{HI}$}&
\multicolumn{1}{c}{$\frac{M_{H_2}}{M_{HI}}$}&
\multicolumn{1}{c}{L$_{H\alpha}$$^c$}&
\multicolumn{1}{c}{$\frac{L_{H\alpha}}{M_{H_2}}$}
\\
       \multicolumn{1}{c}{}&
       \multicolumn{1}{c}{(kpc)}&
\multicolumn{1}{c}{(10$^8$ M$_{\sun}$)}&
       \multicolumn{1}{c}{(cm$^{-2}$)}&
       \multicolumn{1}{c}{(cm$^{-2}$)}&
       \multicolumn{1}{c}{}&
       \multicolumn{1}{c}{(erg s$^{-1}$)}&
       \multicolumn{1}{c}{($\frac{L\sun}{M\sun}$)}\\
\hline
Arp 144 Tail $\#1$&21&$\le$3.3&$\le$4.4 $\times$ 10$^{19}$&
9.7 $\times$ 10$^{19}$&$\le$0.91&&&\\
Arp 144 Tail $\#$2&21&$\le$3.3&$\le$4.4 $\times$ 10$^{19}$&
1.1 $\times$ 10$^{20}$&$\le$0.81&\\
NGC 2814/20 Bridge&5.6&$\le$1.1&$\le$2.0 $\times$ 10$^{20}$\\
MK 208&5.6&$\le$1.1&$\le$1.7 $\times$ 10$^{20}$\\
NGC 3395 Bridge$^d$&5.9&$\le$1.9&$\le$3.1 $\times$ 10$^{20}$\\
NGC 3561 Irregular$^e$&31&$\le$12 &$\le$6.8 $\times$ 10$^{19}$&7 $\times$ 10$^{20}$&$\le$0.20&$\ge$8 $\times$ 10$^{38}$&$\ge$0.00018\\
NGC 3628 Tail $\#1$&3.2&$\le$0.072&$\le$4.4 $\times$ 10$^{19}$\\
NGC 3628 Tail $\#2$&3.2&$\le$0.072&$\le$4.4 $\times$ 10$^{19}$\\
I Zw 192 Plume&43&$\le$62 &$\le$1.8 $\times$ 10$^{20}$\\
SQ A$^f$, 6000 km s$^{-1}$&24&11 &
1.0 $\times$ 10$^{20}$&2.5 $\times$ 10$^{20}$&
0.80&1.2 $\times$ 10$^{41}$&0.016\\
SQ A$^f$, 6700 km s$^{-1}$&24&10&
9.8 $\times$ 10$^{19}$&2.5 $\times$ 10$^{20}$&
0.78&&\\
NGC 7714 Loop&9.9&$\le$2.1&$\le$1.1 $\times$ 10$^{20}$&
9.2 $\times$ 10$^{20}$&$\le$0.24\\
NGC 7714/5 Bridge$^g$&9.9&$\le$0.94&$\le$5.0 $\times$ 10$^{19}$&1.6 $\times$ 10$^{21}$&$\le$0.063&1.8 $\times$ 10$^{39}$&$\ge$0.005\\
NGC 7715 Tail&9.9&$\le$1.0&$\le$5.5 $\times$ 10$^{19}$&
6.5 $\times$ 10$^{20}$&$\le$0.16\\
UGC 12914 Bridge$^d$&15&$\le$190&$\le$4.3 $\times$ 10$^{21}$\\
\\
   \end{tabular}
\end{center}

$^a$All values averaged over the 55$''$ NRAO 12m beam.
$^b$Calculated assuming the standard Galactic 
N$_{H_2}$/I$_{CO}$ 
ratio
(N$_{H_2}$/I$_{CO}$ = 2.8 $\times$ 10$^{20}$ cm$^{-2}$/(K km s$^{-1}$)
and M$_{H_2}$ = 1.1 $\times$ 10$^4$ D$^2$ $\int$S$_V$dV, where
D is the distance in Mpc; Bloemen et al. 1986), 
H$_o$ = 75 km s$^{-1}$ Mpc$^{-1}$, and assuming 34 Jy/K for the 12m
telescope and the source fills the beam
($\eta$$_c$ = 0.82).   As noted in the text, these assumptions
may not hold for all these sources; they are used for comparison
purposes here.
$^c$Not corrected for extinction.
$^d$Because this position is separated from the main disk of the galaxies
by less than half of the FWHM beamwidth, the molecular gas mass is given as
an upper limit here.
$^e$H$\alpha$ luminosity from Duc $\&$ Mirabel 1994.
This was obtained via 
spectroscopy with a 1.5$''$ slit
and so is a lower limit.
$^f$HI values from Shostak et al. 1984.
Both velocity components are included in the L$_{H\alpha}$
and L$_{H\alpha}$/M$_{H_2}$ values given here.
$^g$Derived from the data presented in Smith et al. 1997.

\vfill
\eject

\begin{center}
   {\bf TABLE 6}\\
   PARAMETERS OF ADDITIONAL EXTRA-DISK REGIONS$^a$\\ [12pt]
   \begin{tabular}{crrrrrrrrcccclclcrcccc} \hline
       \multicolumn{1}{c}{Name}&
\multicolumn{1}{c}{$\Theta$}&
\multicolumn{1}{c}{M$_{H_2}$$^b$}&
\multicolumn{1}{c}{N$_{H_2}$$^b$}&
\multicolumn{1}{c}{N$_{HI}$}&
\multicolumn{1}{c}{$\frac{M_{H_2}}{M_{HI}}$}&
\multicolumn{1}{c}{L$_{H\alpha}$$^c$}&
\multicolumn{1}{c}{$\frac{L_{H\alpha}}{M_{H_2}}$}\\
       \multicolumn{1}{c}{}&
       \multicolumn{1}{c}{(kpc)}&
\multicolumn{1}{c}{(10$^8$ M$_{\sun}$)}&
       \multicolumn{1}{c}{(cm$^{-2}$)}&
       \multicolumn{1}{c}{(cm$^{-2}$)}&
       \multicolumn{1}{c}{}&
       \multicolumn{1}{c}{(erg s$^{-1}$)}&
       \multicolumn{1}{c}{($\frac{L\sun}{M\sun}$)}\\
\hline
Arp 143 Tail$^d$&14&$\le$2.0&$\le$5 $\times$ 10$^{19}$&
1.1 $\times$ 10$^{20}$&$\le$0.9\\
Arp 245$^e$&3.5&1.5&3.6 $\times$ 10$^{20}$&2.3 $\times$ 10$^{21}$&
0.31&5.2 $\times$ 10$^{39}$&0.009\\
M81 CO Clump$^f$&0.36&$\sim$0.01&8 $\times$ 10$^{20}$&10$^{21}$&1.6\\
NGC 3077 CO Clump$^g$&0.36&$\sim$0.01&8 $\times$ 10$^{20}$&1.5 $\times$ 10$^{21}$&1.1\\
NGC 2782 East Tail$^h$&14$^d$&6&2 $\times$ 10$^{20}$&
6 $\times$ 10$^{20}$&0.6&4 $\times$ 10$^{39}$&0.002\\
NGC 2782 West Tail$^d$&14&$\le$1.3&$\le$8.5 $\times$ 10$^{19}$
&1.1 $\times$ 10$^{20}$&$\le$0.9\\
NGC 3561 South Tail$^i$&13&2.3&8 $\times$ 10$^{19}$&
2.2 $\times$ 10$^{20}$&0.8&$\ge$1.3 $\times$ 10$^{40}$&$\ge$0.015\\
NGC 4038/9 Tail$^j$&6.1&0.84&1.2 $\times$ 10$^{20}$&
8 $\times$ 10$^{20}$&0.3&1.7 $\times$ 10$^{39}$&0.005\\
NGC 4438 CO Clump$^k$&1.8&8.4&2.0 $\times$ 10$^{21}$&
9.2 $\times$ 10$^{20}$&5&$\le$1.6 $\times$ 10$^{39}$&$\le$0.0005\\
NGC 4410A+B Tail$^l$&26&$\le$8&$\le$4 $\times$ 10$^{19}$&
9 $\times$ 10$^{19}$&$\le$0.9\\
NGC 4676 North Tail$^j$&23&$\le$6.0&$\le$5.9 $\times$ 10$^{19}$&
2.1 $\times$ 10$^{20}$&$\le$0.6&1.1 $\times$ 10$^{40}$&$\ge$0.01\\
NGC 7252 North Tail$^d$&17&$\le$2.4&$\le$4.5 $\times$ 10$^{19}$&1.8 $\times$ 10$^{20}$
&$\le$0.5\\
NGC 7252 South Tail$^d$&17&$\le$4.5&$\le$8.5 $\times$ 10$^{19}$&
1.4 $\times$ 10$^{20}$&$\le$1.2\\

\\
   \end{tabular}
\end{center}

$^a$All values averaged over the 55$''$ NRAO 12m beam, except where
noted.  
$^b$Calculated assuming the standard Galactic 
N$_{H_2}$/I$_{CO}$ 
ratio
(N$_{H_2}$/I$_{CO}$ = 2.8 $\times$ 10$^{20}$ cm$^{-2}$/(K km s$^{-1}$)
and M$_{H_2}$ = 1.1 $\times$ 10$^4$ D$^2$ $\int$S$_V$dV, where
D is the distance in Mpc; Bloemen et al. 1986), 
H$_o$ = 75 km s$^{-1}$ Mpc$^{-1}$, and assuming 34 Jy/K for the 12m
telescope and the source fills the beam
($\eta$$_c$ = 0.82).   As noted in the text, these assumptions
may not hold for all these sources; they are used for comparison
purposes here.
$^c$Not corrected for extinction.
$^d$As tabulated in Smith $\&$ Higdon 1994.
$^e$Using the data from Braine et al. 2000 and Duc et al. 2000, in a
$\sim$23$''$ beam.  The H$\alpha$ luminosity has been corrected for
[N~II] using H$\alpha$/(H$\alpha$+[N~II]) $\sim$ 0.7, as indicated
by the spectroscopy in Duc et al. (2000).
$^f$Using the data from Brouillet et al. 1992 and Rots
$\&$ Shane 1975, in a $\sim$23$''$ beam.
No evidence of star formation at this position has been
observed (Henkel et al. 1993).  The assumed distance is
3.2 Mpc (Sandage $\&$ Tammann 1978; Humphreys $\&$ Aaronson 1987).
$^g$Using the data from Walter $\&$ Heithausen 1999, in
a $\sim$23$''$ beam.  There is
no observed star formation at this position (Walter $\&$ Heithausen 1999).
The HI column density quoted here was estimated from the HI map
in Walter $\&$ Heithausen 2000.
$^h$From data in Smith et al. 1999.  These are averaged over the five
detected positions in the tail, covering an area of $\sim$7645 arcsec$^{-2}$,
the equivalent of an 87$''$ beam.
$^i$Using the data from Braine et al. 2000, Duc $\&$ Mirabel
1994, and Duc et al. 1997, in a $\sim$23$''$ beam.
The H$\alpha$ flux was derived from 
spectroscopy with a 1.5$''$ slit,
and so
is an upper limit.
$^j$The CO measurement is a tentative (4$\sigma$)
detection from Gao et al. 2000; note that
this assumes a broader line (175 km s$^{-1}$)
than the upper limit quoted in Smith $\&$ Higdon
2000.  
The HI measurement is estimated from the Hibbard
et al. (2000) HI map.
The H$\alpha$ measurement is from Mirabel et al. 1992.
$^k$As tabulated in Smith et al. 1999, in a 23$''$ beam.
$^l$Using the data from Smith 2000.

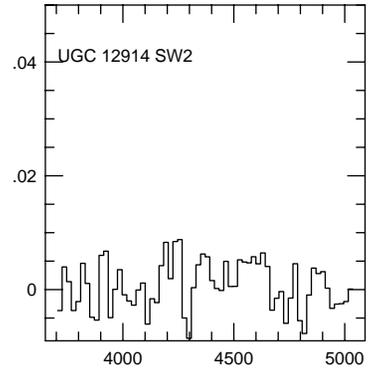
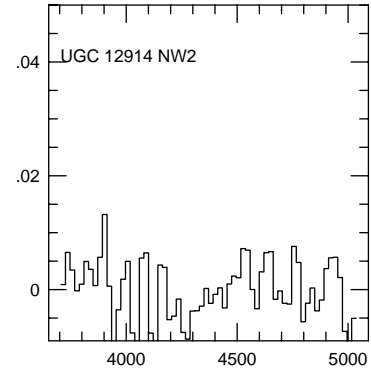
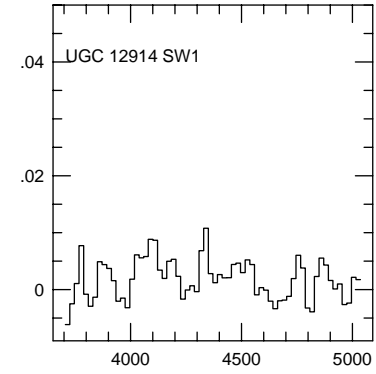
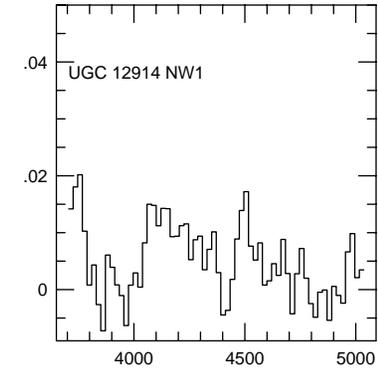
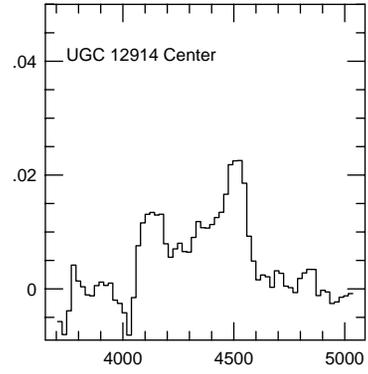
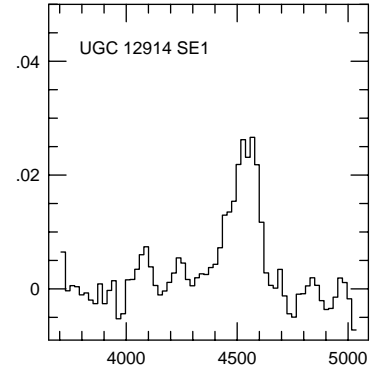
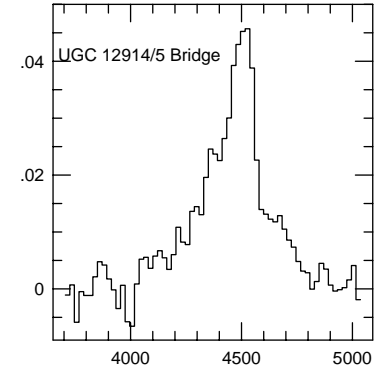
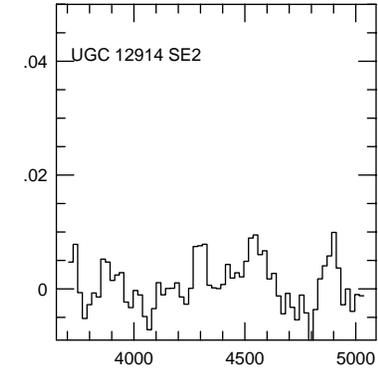
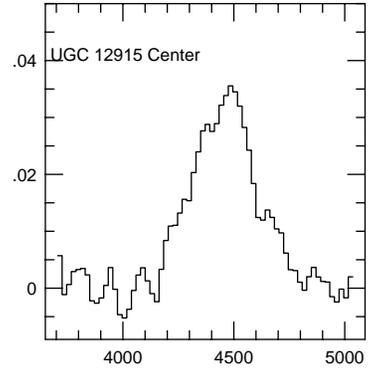
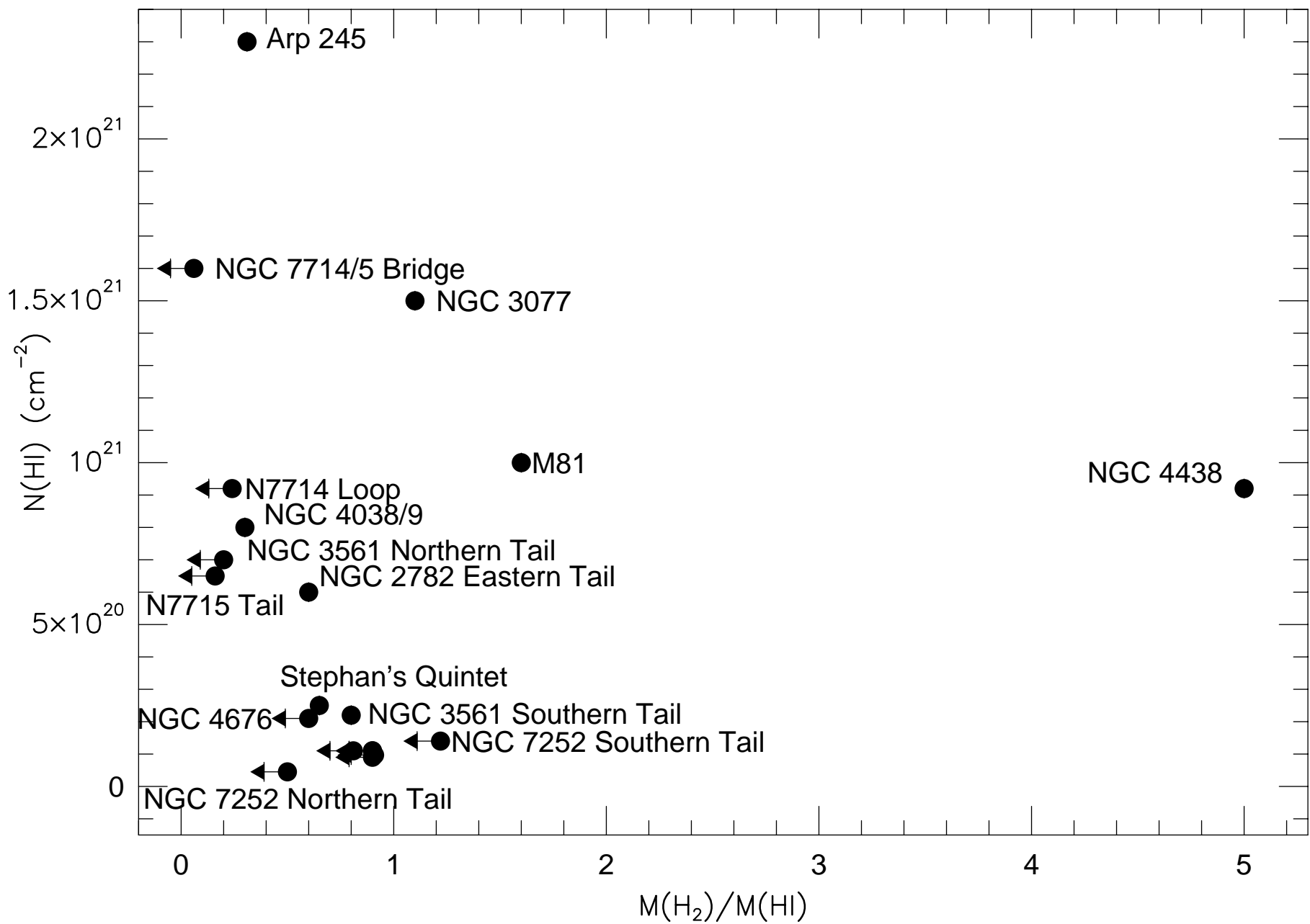
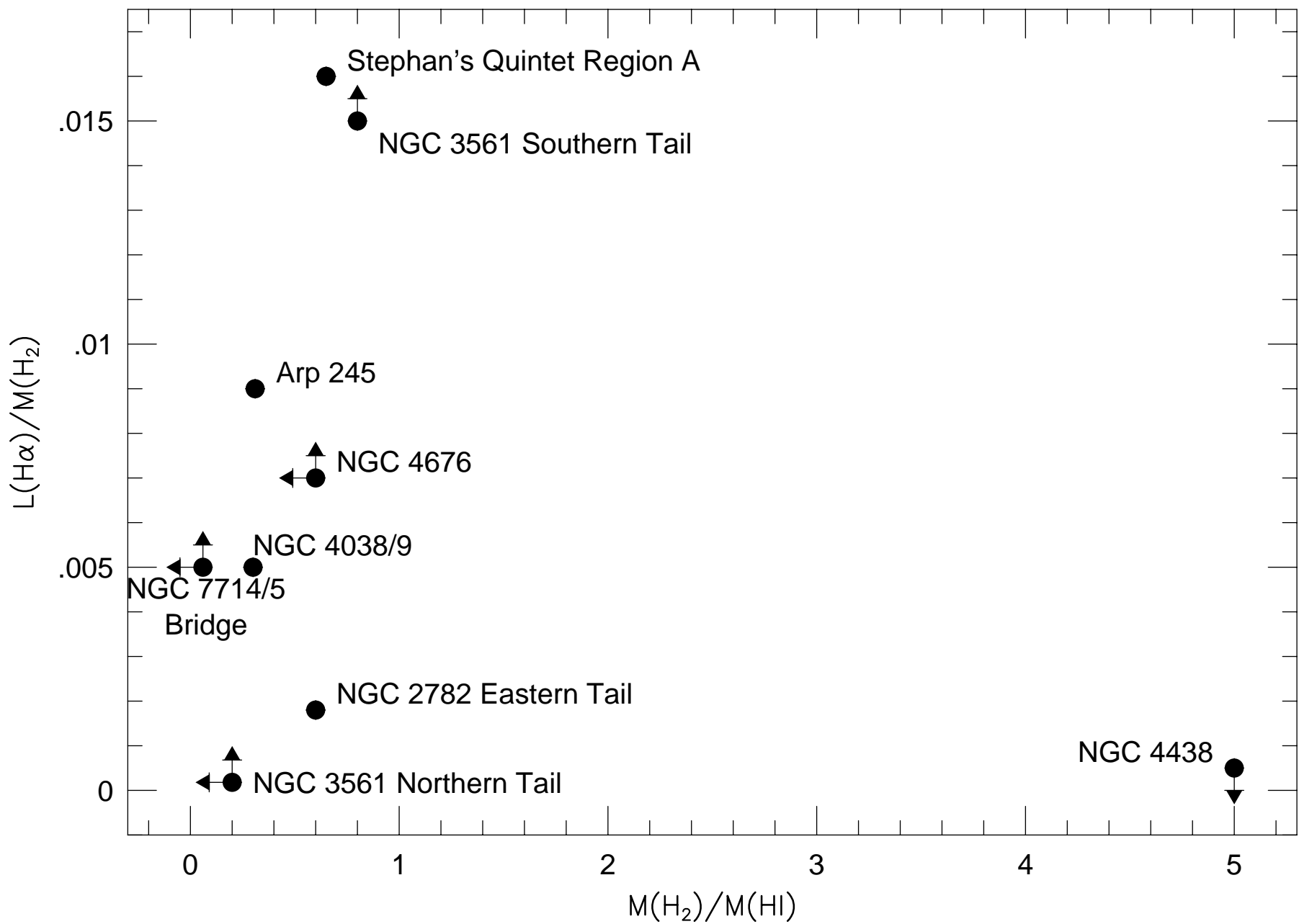
\end{document}